\documentclass[pre,twocolumn,preprintnumbers,amsmath,amssymb]{revtex4}

\usepackage{amsmath,latexsym,epsf,epsfig,graphicx}

\newcommand{\RR}{\mbox{${\mathbb R}$}}

\newcommand{\HH}{\mbox{${\mathbb H}$}}

\def\UU{\mathbb U}

\def\HH{\mathbb H}

\newcommand{\rd}{{\rm d}}

\newcommand{\dQ}{{\dot Q}}
\newcommand{\dS}{{\dot S}}

\newcommand{\Q}{{\mathbb Q}}

\newcommand{\jt}{\widetilde{j}}

\newcommand{\vt}{\vartheta}

\newcommand{\mut}{\widetilde{\mu}}
\newcommand{\nut}{\widetilde{\nu}}

\newcommand{\phl}{\varphi_{{}_L}}
\newcommand{\phr}{\varphi_{{}_R}}

\def\A{\mathcal A} 
\def\H{\mathcal H}

\def\O{\mathcal O}

\def\der{\partial }

\def\ri{{\rm i}}

\def\e{{\rm e}}
\def\eps{\varepsilon}
\def\k{\kappa}

\begin{document}
\title{Anyon Quantum Transport and Noise away from Equilibrium} 

\author{Mihail Mintchev}
\affiliation{
Istituto Nazionale di Fisica Nucleare and Dipartimento di Fisica dell'Universit\`a di Pisa,\\
Largo Pontecorvo 3, 56127 Pisa, Italy}

\author{Paul Sorba} 
\affiliation 
{LAPTh, Laboratoire d'Annecy-le-Vieux de Physique Th\'eorique, 
CNRS, Universit\'e de Savoie,   
BP 110, 74941 Annecy-le-Vieux Cedex, France}
\bigskip 


\begin{abstract}

We investigate the quantum transport of anyons in one space dimension. 
After establishing some universal features of non-equilibrium systems in contact with two 
heat reservoirs in a generalised Gibbs state, we focus on the abelian anyon 
solution of the Tomonaga-Luttinger model possessing axial-vector duality. 
In this context a non-equilibrium representation of the physical observables is constructed, which is the  
basic tool for a systematic study of the anyon particle and heat transport. We determine the associated Lorentz number 
and describe explicitly the deviation from the standard Wiedemann-Franz law induced by the interaction and the 
anyon statistics. The quantum fluctuations generated by the electric and helical currents are investigated and 
the dependence of the relative noise power on the statistical parameter is established.

\end{abstract}

\maketitle

\section{Introduction}

Quantum statistics play a fundamental role in the particle and heat transport in non-equilibrium quantum systems. 
In the present paper we pursue further the study of this feature, focussing on the impact of generalised braid 
statistics \cite{LD-71}-\cite{Wu-84}. More precisely, we consider abelian anyons, 
whose free dynamics in $1+1$ space-time dimensions is described by the Lagrangian density  
\begin{equation} 
{\cal L}_0 = \ri \psi_1^*(\der_t - \der_x)\psi_1 +  \ri \psi_2^*(\der_t + \der_x)\psi_2 \, .  
\label{lagrangian0}
\end{equation} 
Here $\{\psi_s (t,x)\, :\, s=1,2\}$ are complex fields obeying {\it anyon} equal-time exchange relations \cite{LMR-95} ($x_1\not=x_2$)
\begin{equation}
\psi_s^*(t,x_1) \psi_s (t,x_2) = 
\e^{(-1)^{s}\, \ri \pi \, \k\, \eps(x_1-x_2)} \psi_s (t,x_2)\psi_s^*(t,x_1)\, , 
\label{exch}
\end{equation} 
where $\eps(x)$ is the sign function and $\k > 0$ is the so called 
{\it statistical parameter}, which interpolates between bosons ($\k$ - even integer) and 
fermions ($\k$ - odd integer). The parameter $\k$ plays a central role in our investigation, being 
devoted to a systematic study of the $\k$-dependence of the anyon quantum transport 
and the noise generated away from equilibrium. 

The interaction, which successfully describes \cite{H-81, Hprl-81} the universal features of a large class of one-dimensional 
systems exhibiting gapless excitations with linear spectrum, is fixed by 
\begin{equation} 
{\cal L}_I = -\frac{\pi g_+}{2}(\psi_1^* \psi_1+\psi_2^* \psi_2)^2 - \frac{\pi g_-}{2}(\psi_1^* \psi_1-\psi_2^* \psi_2)^2\, ,   
\label{lagrangianI}
\end{equation} 
where $g_\pm \in \RR$ are the coupling constants. The use of the normalisation factor $\pi$ in (\ref{lagrangianI}) 
simplifies some basic equations in what follows and is introduced for convenience.  

The total Lagrangian ${\cal L}_0 +{\cal L}_I$ does not involve dimensional parameters and is 
{\it scale invariant}. Combined with (\ref{exch}) it defines the dynamics of anyon \cite{LMP-00, IT-01} 
Tomonaga-Luttinger (TL) liquid \cite{T-50}-\cite{L-63} with 
$\UU(1)\times {\widetilde \UU}(1)$ symmetry, where the $\UU(1)$-vector  
and ${\widetilde \UU}(1)$-axial transformations are defined by 
\begin{equation}
\psi_s (t,x) \longmapsto \e^{\ri \alpha} \psi_s(t,x)\, , \quad \alpha \in [0,2\pi)\, , 
\label{e1}
\end{equation} 
and  
\begin{equation}
\psi_s (t,x) \longmapsto \e^{\ri (-1)^s {\widetilde \alpha}} \psi_s(t,x)\, , \quad  {\widetilde \alpha} \in [0,2\pi)\, , 
\label{chi1}
\end{equation} 
respectively. The relative conserved currents describe the electric and helical transport respectively. 

\begin{figure}[h]
\begin{center}
\begin{picture}(600,30)(116,270) 
\includegraphics[scale=0.75]{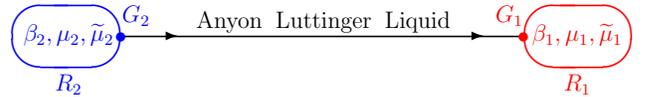}
\end{picture} 
\end{center}
\caption{(Color online) Anyon Luttinger liquid connected to two GGE reservoirs $R_1$ and $R_2$.} 
\label{fig1}
\end{figure}

The main subject of our investigation is the quantum transport induced by connecting the system ${\cal L}_0 +{\cal L}_I$
via the gates $G_i$ with two heat reservoirs $R_i$ as shown in Fig. \ref{fig1}. Each of them 
is described by a Generalised Gibbs Ensemble (GGE) with (inverse) temperatures $\{\beta_i\geq 0\, :\, i=1,2\}$ and the 
chemical potentials $\{\mu_i, \mut_i \, :\, i=1,2\}$, where $\mu$ and $\mut$ are associated with the commuting charges $\Q$ and 
$\widetilde \Q$ generating the $\UU(1)\times {\widetilde \UU}(1)$ symmetry. 
The interaction of the anyons emitted and absorbed by the two heat reservoirs $R_i$ is 
described by (\ref{lagrangianI}). This interaction drives 
the system away from equilibrium. The capacity of $R_i$ is assumed large enough so that the emission-absorption 
processes do not change their parameters. 

Systems with the structure shown in Fig. \ref{fig1} in one space dimension are subject of intense 
studies. They are successfully applied for describing the transport properties of quantum 
wire junctions \cite{kf-92}-\cite{KGA-20} and fractional quantum Hall samples,  
where abelian anyons with filling factors $\k = 1/(2n+1)$ with $n=1,2,...$ are propagated \cite{AS-08}. 
Recently there is also the exciting possibility to conceive in laboratory 
the setting in Fig. \ref{fig1} by ultracold Bose gases \cite{BDZ-08}-\cite{CS-16}. 
The remarkable control over the interactions and the geometry of the samples in such experiments 
allow to explore a variety of fundamental aspects of many-body quantum physics. This is also the case for the anyon 
systems considered in this paper. 

At the theoretical side the equilibrium fermionic version 
of the TL model with $\k=1$ is widely investigated \cite{ML-65}-\cite{Egg-08} in the literature. 
There has been also great interest \cite{GGM-08}-\cite{MS-13} in the behaviour of the model away from equilibrium. These 
more recent studies focus essentially on the charge transport induced by the difference $\mu_1-\mu_2$. 
However, the combined effect of the two chemical potentials $\mu$ and $\mut$ with generic statistical parameter $\k$ 
is poorly studied. Filling this gap is among our main goals below. In this respect we demonstrate 
that the presence of both $\mu$ and $\mut$ makes explicit the {\it axial-vector duality}  
that characterises the TL model (\ref{lagrangian0},\ref{lagrangianI}). Let us recall in this respect that 
this duality is broken in the presence of energy preserving impurities, which necessarily violate \cite{MS-13} at least 
one among the factors of the $\UU(1)\times {\widetilde \UU}(1)$ symmetry. We show that the non-trivial 
interplay between $\mu$ and $\mut$ generates {\it persistent} charge and heat currents at equilibrium. 
We also establish the relevant impact of $\mu$ and $\mut$ on the entropy 
production in the system away from equilibrium.  

Another related aspect, addressed in the paper, concerns the  quantum fluctuations generated by the 
electric and helical currents. Such fluctuations produce noise which spoils the propagation of the anyon  
excitations. It is known \cite{L-98}-\cite{SB-06} however that the current fluctuations carry also useful information, 
providing the experimental basis of noise spectroscopy. In this respect we exhibit the dependence 
of the noise on the statistical parameter $\k$ in explicit form. 

The paper is organised as follows. In the next section we focus on the 
universal properties of non-equilibrium quantum systems in contact with two GGE heat reservoirs. 
In section 3 we summarise the operator solution of the TL model (\ref{lagrangian0},\ref{lagrangianI}) 
and construct in detail a specific representation of this solution, which 
describes the system connected with the two heat reservoirs. On this basis we investigate in section 4 
the anyon particle and heat transport and evaluate the associated Lorentz number. 
The impact of the interaction $g_\pm$ and the statistical parameter $\k$ on the 
Wiedemann-Franz law is a central point of this section. We investigate here also 
the mean value of the entropy production in the presence of 
all chemical potentials characterising the GGE. Section 5 is devoted to the quantum 
fluctuations of the electric and helical currents and the impact of $\k$ on the generated noise. 
Finally, section 6 collects our conclusions.

\section{Basic features of quantum transport between two GGE reservoirs} 

Before discussing the specific anyon case, we would like to establish 
some general properties of the quantum transport between two GGE reservoirs 
as shown in Fig. \ref{fig1}. For simplicity we assume in what follows that the total energy 
$E$ and the charges $\{\Q\, ,{\widetilde \Q}\}$ are conserved, which means that 
the incoming and outgoing energy and charge flows through the gates $G_i$ compensate each other.  
We show below that under this realistic physical assumption, the system posses 
remarkable universal features, which do not depend on the nature of the interaction 
and are therefore of great relevance. In order to establish them, we introduce  
the energy and charge densities $\vt_t$ and $j_t,\, {\tilde j}_t$, satisfying  
\begin{eqnarray}
&\HH& = \int_{G_2}^{G_1} \rd x\, \vt_t(t,x)\, , 
\label{g1a}\\
&\Q& = \int_{G_2}^{G_1} \rd x\, j_t(t,x)\, , \quad {\widetilde \Q} = \int_{G_2}^{G_1} \rd x\, {\widetilde j}_t(t,x)\, .
\label{g1b}
\end{eqnarray} 
Let $\vt_x$ and $j_x,\, {\widetilde j}_x$ be the corresponding local conserved currents, which obey the continuity equations 
\begin{eqnarray}
&\der_t \vt_t(t,x)& - \, \der_x \vt_x(t,x)= 0\, ,
\label{g2} \\
&\der_t j_t(t,x)& - \, \der_x j_x(t,x) = \der_t {\widetilde j}_t(t,x) - \, \der_x {\widetilde j}_x(t,x) = 0\, . 
\label{g3} 
\end{eqnarray} 
Combining (\ref{g1a},\ref{g1b}) with (\ref{g2},\ref{g3}) one finds 
\begin{eqnarray}
{\dot \HH}= 0\quad  &\Rightarrow& \quad \vt_x(t,G_1) = \vt_x(t,G_2)\, ,
\label{g4} \\
{\dot \Q} = 0\quad  &\Rightarrow& \quad j_x(t,G_1) = j_x(t,G_2)\, , 
\label{g5}\\  
{\dot {\widetilde \Q}} = 0\quad  &\Rightarrow& \quad {\widetilde j}_x(t,G_1) = {\widetilde j}_x(t,G_2)\, , 
\label{g5t}
\end{eqnarray} 
which is the expected behaviour in the gates $G_i$ for any $t$. 

Let us consider now the heat current $q_x$ flowing through $G_i$. Since the values of the chemical potentials 
in $G_i$ are $\mu_i$ and $\mut_i$, one has, following the rules of non-equilibrium thermodynamics \cite{Callen},   
\begin{equation}
q_x(t,G_i) = \vt_x (t,G_i) - \mu_i j_x(t,G_i)- \mut_i {\widetilde j}_x(t,G_i)\, .  
\label{g6} 
\end{equation} 
From (\ref{g6}) it follows that for $\mu_1\not = \mu_2$ and/or  $\mut_1\not = \mut_2$ the 
heat flow through $G_1$ differs from that through 
$G_2$. In fact, 
\begin{eqnarray}
\dQ \equiv q_x(t,G_1)-q_x(t,G_2) = \qquad \qquad \qquad \quad 
\nonumber \\
(\mu_2-\mu_1) j_x(t,G_1) + (\mut_2-\mut_1) {\widetilde j}_x(t,G_1)\not=0\, . 
\label{g7}
\end{eqnarray}
At this point we use that the total energy of the system has $3$ components - {\it heat} energy and $2$ different types of 
{\it chemical} energies associated with the charges $\Q$ and $\widetilde \Q$. 
Since the total energy is conserved, (\ref{g7}) implies that heat energy can be converted in one or two types of chemical energies and 
vice versa. This process depends on the state $\Psi \in \H$ of the system, namely on 
the expectation value 
\begin{equation}
\langle \dQ \rangle_{{}_\Psi} \equiv (\Psi\, ,\, \dQ \Psi) \, , 
\label{g8}
\end{equation}
where $(\cdot\, ,\, \cdot )$ is the scalar product in the state space $\H$. Chemical energy is converted to heat energy 
if $\langle \dQ \rangle_{{}_\Psi} >0$. The opposite process takes place for $\langle \dQ \rangle_{{}_\Psi} <0$ and energy 
transmutation is absent only if $\langle \dQ \rangle_{{}_\Psi} =0$. It is worth stressing that there is no dissipation 
in the energy conversion. 

The above argument uses only symmetry considerations and does not involve the dynamics. 
Depending explicitly on the heat flow, this phenomenon of energy conversion \cite{MSS-15} has a 
relevant impact on the heat transport in the system in Fig. \ref{fig1}. In fact, evaluating the mean value of (\ref{g7}) 
in the state $\Psi$, one has 
\begin{equation}
\langle q_x(t,G_1) \rangle_{{}_\Psi}  = \langle q_x(t,G_2) \rangle_{{}_\Psi} +\langle \dQ \rangle_{{}_\Psi} \, . 
\label{g9}
\end{equation}
Therefore one concludes that the observed mean values of the heat flow through 
the gates $G_1$ and $G_2$ are in general different. 

Notice that we adopt here only the value of the heat current in the gates $G_i$. The point is that the 
heat current in the interaction domain $\RR$ in Fig. \ref{fig1} is not known, because the temperature 
and the chemical potentials 
are not determined in this region. In order to introduce the concept of local 
temperature $\beta(x)$ and chemical potentials $\mu(x)$ 
and $\mut(x)$ in a point $x \in \RR$ 
one needs further model dependent assumptions \cite{LLMM-17}, which are not needed for our construction.  

Let us discuss now the choice of the state $\Psi$. In this paper we consider steady states which are 
generated by the GGE states of the heat reservoirs and are 
invariant under time translations, implying that the expectation value $\langle \O(t,x) \rangle_{{}_\Psi}$ 
of any observable $\O$ is actually $t$-independent. In particular, even if $\dQ$ depends on $t$, its 
expectation value $\langle \dQ \rangle_{{}_\Psi}$ does not. 

Concerning the action of the time reversal operation on $\Psi$, we first observe that for 
steady states, describing the non-equilibrium system 
in Fig. \ref{fig1}, there is a nontrivial energy exchange between the reservoirs $R_i$ leading to 
\begin{equation}
\langle \vt_x(t,x) \rangle_{{}_\Psi} \equiv (\Psi\, ,\, \vt_x(t,x) \Psi) \not = 0\, . 
\label{g10}
\end{equation}
Now we recall that \cite{Weinberg}
\begin{equation}
 T\, \vt_x(t,x)\, T^{-1} = -\vt_x(-t,x)\, ,  
\label{g11}
\end{equation}
where $T$ is the anti-unitary operator implementing the time reversal in the Hilbert space $\H$. 
Taking the expectation value of (\ref{g11}) one has 
\begin{equation}
\langle T\, \vt_x(t,x)\, T^{-1}\rangle_{{}_\Psi} = -\langle \vt_x(-t,x)\rangle_{{}_\Psi}\, ,  
\label{g12}
\end{equation}
which, combined with the fact that $\langle \vt_x(t,x)\rangle_{{}_\Psi}$ is $t$-inde\-pendent, implies that 
\begin{equation}
T\, \Psi \not = \Psi \, . 
\label{g13}
\end{equation} 
Therefore time reversal is spontaneously broken in any state $\Psi$ obeying (\ref{g10}). 
This genuine quantum field theory phenomenon is the origin of the non-trivial entropy production 
\begin{equation}
\langle \dS \rangle_{{}_\Psi} =  \beta_1 \langle q_x(t,G_1) \rangle_{{}_\Psi} 
-\beta_2 \langle q_x(t,G_2) \rangle_{{}_\Psi}\, , 
\label{g14}
\end{equation}
which takes place even for systems with time reversal invariant dynamics without dissipation. 

Summarising, the physical consequences of the energy and charge conservation in systems, 
schematically represented in Fig. \ref{fig1}, are: 

(i) conversion of heat to chemical energy or vice versa; 

(ii) non-trivial entropy production. 

These features do not depend on the dynamics being therefore universal. 
What follows is an illustration of (i) and (ii) and their impact on the particle and heat 
transport in the anyon TL model defined by (\ref{lagrangian0},\ref{lagrangianI}). 

\section{Anyon Luttinger liquid} 

In this section, following \cite{LMP-00, bcm-09}, we first briefly summarise the anyon operator solution of the TL model  
(\ref{lagrangian0},\ref{lagrangianI}). Afterwards we provide a new Hilbert space representation of this solution, 
which is induced by the non-equilibrium steady state describing the system connected with the 
two GGE reservoirs as shown in Fig. \ref{fig1}.

\subsection{Operator solution} 

The classical equations of motion of the TL model read 
\begin{eqnarray}
&& \ri (\der_t -\der_x) \psi_1=  \pi g_+ j_t \psi_1  
+  \pi g_- \jt_t \psi_1\, , 
\label{eqm1}\\
&& \ri (\der_t +\der_x) \psi_2 =  \pi g_+ j_t \psi_2  
- \pi g_- \jt_t \psi_2    \, ,
\label{eqm2}
\end{eqnarray}
where 
\begin{equation}
j_t= \left (\psi_1^*\psi_1 + \psi_2^*\psi_2 \right )\, , 
\quad \jt_t= \left (\psi_1^*\psi_1 - \psi_2^*\psi_2 \right )\, ,
\label{chargedensities}
\end{equation}
are the charge densities generating the $\UU(1)\otimes {\widetilde \UU}(1)$ conserved charges. 
It is well known that the operator solution of (\ref{eqm1},\ref{eqm2}) is obtained 
via bosonisation (see e.g. \cite{H-81}) in terms of the chiral scalar fields 
$\phr$ and $\phl$ satisfying 
\begin{equation}
(\der_t + v\der_x)\phr(vt-x) = 0\, , \quad (\der_t - v\der_x)\phl (vt+x) = 0\, , 
\label{LR}
\end{equation} 
where $v$ is a velocity specified later on. Referring for the details to \cite{bcm-09}, 
the anyon operator solution is given by 
\begin{eqnarray}  
\psi_1(t,x) &=&
\eta :\e^{\ri \sqrt {\pi} \left [\sigma \phr (vt-x) + \tau \phl (vt+x)\right ]}:\, , 
\label{psi1}\\
\psi_2(t,x) &=&
\eta :\e^{\ri \sqrt {\pi} \left [\tau \phr (vt-x) + \sigma \phl (vt+x)\right ]}:\, . 
\label{psi2}
\end{eqnarray}
Here $\eta$ is a Klein factor, whose explicit form is not relevant for what follows.  
The parameters $\sigma,\, \tau \in \RR$ are determined below and 
$: \cdots :$ denotes the normal product in the algebra of the chiral fields (\ref{LR}). 
Inserting (\ref{psi1},\ref{psi2}) in the anyon exchange relation (\ref{exch}) one finds 
\begin{equation}
\zeta_+ \zeta_- = \k \, , \qquad \zeta_\pm =\tau\pm\sigma \, .  
\label{zpm}
\end{equation} 
For the charge densities and relative currents one has 
\begin{eqnarray}
j_t(t,x)  = 
\frac{-1}{2\sqrt {\pi }\zeta_+} 
\left [(\der \phr)(vt-x) + (\der \phl)(vt+x)\right ] , 
\label{jt} \\ 
\jt_t(t,x)  = 
\frac{-1}{2\sqrt {\pi }\zeta_-} 
\left [(\der \phr)(vt-x) - (\der \phl)(vt+x)\right ] , 
\label{tjt} 
\end{eqnarray} 
and 
\begin{eqnarray}
j_x(t,x) = 
\frac{v}{2\sqrt {\pi }\zeta_+} 
\left [(\der \phr)(vt-x) - (\der \phl)(vt+x)\right ]\, , 
\label{jx}\\ 
\jt_x(t,x) = 
\frac{v}{2\sqrt {\pi }\zeta_-} 
\left [(\der \phr)(vt-x) + (\der \phl)(vt+x)\right ]\, . 
\label{tjx}
\end{eqnarray} 
Because of (\ref{LR}), these densities and currents satisfy the conservation law (\ref{g3}).  
Plugging (\ref{jt},\ref{tjt}) in the equations of motion (\ref{eqm1},\ref{eqm2}) and using (\ref{zpm}) one finds finally 
\begin{eqnarray}
\zeta_\pm^2 &=& \kappa
\left(\frac{\kappa +g_+}{\kappa +g_-}\right)^{\pm 1/2}\, , 
\label{z}\\ 
v&=&\frac{1}{\kappa}\sqrt{(\kappa +g_-)(\kappa +g_+)}\, , 
\label{v}
\end{eqnarray} 
where the positive roots are taken in the right hand side. Equations (\ref{z},\ref{v}) 
determine the parameters $\sigma$ and $\tau$ and the velocity $v$ of the interacting 
anyons in terms of the coupling constants $g_\pm$ and the statistical parameter $\k$. 
Notice that the velocity $v$ of the interacting anyons differs from the free velocity $v_0$ (according to our conventions 
(\ref{lagrangian0}) $|v_0|=1$)  
and depends on $\k$ and $g_\pm$ as well. 
We assume in what follows that $\{\k,\, g_\pm\}$, defining the abelian anyon Luttinger liquid, belong to the domain  
\begin{equation}
{\cal D} =\{\k>0,\; \kappa > -g_\pm\} \, ,  
\label{physcond}
\end{equation} 
which ensures that $\sigma$, $\tau$ and $v$ are real and finite. 

In conclusion, we observe that the above anyon solution of the TL model for generic $\k>0$ 
generates for $\k=1$ the fermionic solution, usually described in the literature \cite{ML-65}-\cite{Egg-08} 
in terms of the parameters $\{g_2,\, g_4,\, K\}$ related to $\{g_+,\, g_-,\, \zeta_\pm\}$ as follows: 
\begin{eqnarray}
g_2 &=& \frac{1}{2}(g_+-g_-)\, , \quad g_4 = \frac{1}{2}(g_+ + g_-)\, , 
\label{correspondence1}\\
\quad K &=& \zeta^2_-{}_{\vert_{\k=1}}= \zeta^{-2}_+{}_{\vert_{\k=1}}\, . 
\label{correspondence2} 
\end{eqnarray}

\subsection{Representation implementing the GGE reservoirs}

Our goal now is to construct a representation of the chiral fields (\ref{LR}), which implements the GGE 
reservoirs $R_i$ in the operator solution (\ref{psi1}-\ref{tjx}). For this purpose we use the standard decomposition 
\begin{eqnarray}
\phr (\xi) = 
\int_0^\infty \rd k \frac{\sqrt {\Delta(k)}}{\pi \sqrt 2}\left [a^\ast (k) \e^{ik\xi} + 
a(k)\e^{-ik\xi} \right ] \, ,
\label{phr} \\
\phl (\xi) =
\int^0_{-\infty } \rd k \frac{\sqrt {\Delta(k)}}{\pi \sqrt 2} \left [a^\ast (k) \e^{-ik\xi} + 
a(k) \e^{ik\xi}\right ]\, , 
\label{phl}
\end{eqnarray} 
where 
\begin{equation} 
|k| \Delta(k) = 1\, , 
\label{modk}
\end{equation} 
and choose suitable representations of the two canonical commutation 
algebras $\A_\pm = \{a(k),\, a^*(k)\, :\, k\gtrless 0 \}$. 
Since the origin of the right moving field (\ref{phr}) is the reservoir $R_1$, we take for $\A_+$ the Gibbs 
representation at temperature $\beta_1$. For analogous reason we adopt $\A_-$ in the Gibbs representation 
with temperature $\beta_2$. More explicitly, consider the Bose distribution 
\begin{equation} 
d_i (k) = \frac{\e^{-\beta_i [|k| - \lambda]}}{1- \e^{-\beta_i [|k| - \lambda]}} \, , \quad \lambda <0\, , \quad i=1,2\, . 
\label{bd} 
\end{equation} 
Then 
\begin{equation}
\langle a^*(p)a(k) \rangle = 
\begin{cases}
d_1(k)\, 2\pi \delta (k-p)\, , &\quad  p,\, k > 0 \, , \\
d_2(k)\, 2\pi \delta (k-p)\, , &\quad  p,\, k < 0 \, . 
\end{cases} 
\label{bd1}
\end{equation}
The bosonic chemical potential $\lambda<0$ allows to avoid the infrared singularity at $k=0$ in (\ref{bd}).  
The limit $\lambda \to 0^-$ exists \cite{LMP-00} and is performed in the correlation functions of the TL observables. 

Once we have the temperatures $\beta_i$ via the Gibbs representation of $\A_\pm$, we have to introduce the chemical 
potentials $\mu_i$ and $\mut_i$. At this point we generalise away from equilibrium the strategy of \cite{LMP-00}, performing the shifts 
\begin{eqnarray}
\phr (\xi) \mapsto \phr (\xi) - \frac{1}{v\sqrt \pi}\left (\frac{\mu_1}{\zeta_+} + \frac{\mut_1}{\zeta_-}\right ) \xi \, , 
\label{phrs}\\
\phl (\xi) \mapsto \phl (\xi) - \frac{1}{v\sqrt \pi}\left (\frac{\mu_2}{\zeta_+} - \frac{\mut_2}{\zeta_-}\right ) \xi \, ,
\label{phls}
\end{eqnarray}
where $\zeta_\pm$ and $v$ are given by (\ref{z},\ref{v}). 
The form of (\ref{phrs},\ref{phls}) respects the equations of motion (\ref{LR}) and is fixed by requiring 
that at equilibrium 
\begin{equation}
\beta_1=\beta_2\equiv \beta\, , \quad \mu_1=\mu_2\equiv \mu\, , \quad \mut_1=\mut_2 \equiv \mut\, , 
\label{eq}
\end{equation}
the correlation functions of $\psi_s$ satisfy the Kubo-Martin-Schwinger (KMS) condition \cite{Haag}. 
Since the latter is a basic condition in our construction, it is instructive to discuss the issue in detail. 

Introducing the notation $t_{12}\equiv t_1-t_2\, ,\; x_{12}\equiv x_1-x_2$ and taking into account that 
$\langle \eta^* \eta \rangle =1$ one finds in the limit $\lambda_b \to 0^-$
\begin{eqnarray}
\langle \psi_1^*(t_1,x_1)\psi_1(t_2,x_2)\rangle = 
\e^{-\ri F(t_{12},x_{12})}
\nonumber \\
\times \left [\frac{\beta_1}{\pi} \sinh \left (\frac{\pi}{\beta_1}(vt_{12}- x_{12}) -
\ri \varepsilon \right )\right ]^{-\tau^2}  \; \; \; 
\nonumber \\
\times \left [\frac{\beta_2}{\pi} \sinh \left (\frac{\pi}{\beta_2}(vt_{12} + x_{12}) -
\ri \varepsilon \right )\right ]^{-\sigma^2},  
\nonumber \\
\label{corr11}
\end{eqnarray} 
where $\varepsilon \to 0^+$ and the phase factor $F$ is given by 
\begin{eqnarray}
F(t,x) = \qquad \qquad \qquad \qquad \qquad
\nonumber \\
-\frac{\sigma}{v}\left (\frac{\mu_1}{\zeta_+} + \frac{\mut_1}{\zeta_-}\right )(vt-x)  
- \frac{\tau}{v}\left (\frac{\mu_2}{\zeta_+} - \frac{\mut_2}{\zeta_-}\right )(vt+x) .  
\label{phase}
\end{eqnarray} 
The correlation function 
$\langle \psi_2^*(t_1,x_1)\psi_2(t_2,x_2)\rangle$ is obtained by 
performing $\sigma \leftrightarrow \tau$ in (\ref{corr11}). 

At equilibrium (\ref{eq}) one infers from (\ref{corr11}) 
\begin{eqnarray}
\langle \psi_1^*(t_1,x_1)\psi_1(t_2,x_2)\rangle_{\rm eq} = 
\e^{-\ri F_{\rm eq}(t_{12},x_{12})}
\nonumber \\
\times \left [\frac{\beta }{\pi} \sinh \left (\frac{\pi}{\beta }(vt_{12}- x_{12}) -
\ri \varepsilon \right )\right ]^{-\tau^2}  \; 
\nonumber \\
\times \left [\frac{\beta }{\pi} \sinh \left (\frac{\pi}{\beta }(vt_{12} + x_{12}) -
\ri \varepsilon \right )\right ]^{-\sigma^2}  
\nonumber \\
\label{corr11eq}
\end{eqnarray} 
with 
\begin{equation}
F_{\rm eq}(t,x) = -\mu t + \mut t - \frac{x}{v} \left (\frac{\zeta_-}{\zeta_+} \mu - \frac{\zeta_+}{\zeta_-}\mut \right ) \, . 
\label{phaseeq}
\end{equation}
On the other hand, the KMS automorphism acts on $\psi_s$ as follows \cite{Haag}
\begin{equation}
\gamma_\alpha \; :\; \psi_s(t,x) \longmapsto \psi_s (t+\alpha )\, \e^{\ri \alpha [\mu + (-1)^s \mut ]} \, , \quad \alpha \in \RR\, . 
\label{kms1}
\end{equation}
Now one can directly verify that the KMS condition 
\begin{eqnarray}
\langle \psi_s^*(t_1,x_1)[\gamma_{\left (\alpha +\ri \frac{\beta}{v}\right )} \psi_s](t_2,x_2)\rangle = 
\nonumber \\
\langle [\gamma_{\alpha} \psi_s](t_2,x_2)\psi_s^*(t_1,x_1)\rangle \qquad  
\label{kms2}
\end{eqnarray}
holds at equilibrium, which concludes the argument. 

Let us comment in conclusion on scale invariance. 
Although the dynamics, associated with the TL Lagrangian (\ref{lagrangian0},\ref{lagrangianI}) is scale invariant, 
the correlation functions of $\psi_s$ violate this symmetry because they 
are computed in a non-equilibrium state, which depends on parameters 
like temperatures and chemical potentials. The scale invariant limit is obtained 
by $\mu_i \to 0$, $\mut_i \to 0$ and $\beta_i \to \infty$. Performing this limit 
in (\ref{corr11},\ref{phase}) one finds the following 
scaling dimension for $\psi_s$ 
\begin{equation}
d_\psi = \frac{1}{2}(\tau^2 + \sigma^2) = \frac{\k (2 \k +g_+ + g_-)}{4\sqrt{(\kappa +g_-)(\kappa +g_+)}}\, , 
\label{scdim}
\end{equation}
showing a relevant modification of the canonical dimension $1/2$ of $\psi_s$ due to both dynamics and statistics. 

Summarising, we presented the operator solution of the TL model in terms of the chiral scalar fields 
$\{\phr\, ,\phl\}$ and constructed a representation of these fields implementing the GGE reservoirs $R_i$ 
at different temperatures and chemical potentials. This representation obeys in the equilibrium limit the KMS condition 
and respects the conservation of the total 
energy and the charges $\Q$ and ${\widetilde \Q}$. At this stage we are therefore ready to 
investigate the quantum transport in the above non-equilibrium state and test 
the universal features, pointed out in the previous section. 

\section{Anyon quantum transport}

In this section we derive and study the mean values of the charge and heat currents 
flowing in the system shown in Fig. \ref{fig1}. The invariance under space-time translations implies 
that these mean values are both $x$ and $t$-independent.

\subsection{Electric and helical transport} 

Taking into account (\ref{phrs},\ref{phls}), one gets for the expectation value of the electric (vector) current (\ref{jx}) 
\begin{equation}
\langle j_x \rangle = 
\frac{\mu_2-\mu_1}{2\pi \zeta_+^2}  - \frac{\mut_1+\mut_2}{2\pi \zeta_+ \zeta_-} \, . 
\label{t1}
\end{equation}
By means of (\ref{zpm},\ref{z}) one can reconstruct the explicit dependence on the statistical parameter $\k$. One finds 
\begin{equation}
\langle j_x \rangle = \frac{1}{2\pi \k} \left [(\mu_2-\mu_1) \sqrt{\frac{\k  +g_-}{\k  +g_+}} - (\mut_1+\mut_2) 
\right ] \, ,   
\label{t2}
\end{equation} 
which reveals a remarkable feature of the chiral chemical potentials $\mut_i$ to drive a non-vanishing 
electric current even for $\mu_1=\mu_2$. At equilibrium (\ref{eq}) one has 
\begin{equation}
\langle j_x \rangle_{\rm eq} = -\frac{\mut }{\pi \, \k} \, ,   
\label{t3}
\end{equation} 
which precisely coincides with the persistent current discovered in \cite{LMP-00} and proportional to $\k^{-1}$. 
This current has a simple physical origin. The point is that at equilibrium the chemical potential 
$\mut$ can be equivalently implemented \cite{LMP-00} by coupling the TL model (\ref{lagrangian0},\ref{lagrangianI}) 
with a constant magnetic field $h=\mut$. 

An expression similar to (\ref{t2}) holds also for the helical (axial) current 
\begin{equation}
\langle \jt_x \rangle = \frac{1}{2\pi \k} \left [(\mut_2-\mut_1) \sqrt{\frac{\k  +g_+}{\k  +g_-}} - (\mu_1+\mu_2) 
\right ] \, . 
\label{t4}
\end{equation}
It is worth mentioning that (\ref{t2}) and (\ref{t4}) are related by the transformation 
\begin{equation}
\mu_i \leftrightarrow \mut_i\, , \qquad g_+ \leftrightarrow g_- \, , 
\label{duality}
\end{equation}
which implements the axial-vector duality in the model and 
confirms the deep interplay between helical and electric transport in 
the Luttinger liquid. We observe also that the currents (\ref{t1},\ref{t4}) depend 
on the chemical potentials, but not on the temperatures. Therefore there is no thermo-electric effect at the level 
of {\it mean values}, which agrees with the general prediction \cite{BD-15, HL-18} from non-equilibrium 
conformal field theory (CFT). We stress however that the {\it quantum fluctuations} of these currents, 
derived in section 5 below, are instead temperature depend.

\subsection{Heat transport}

In the basis of the chiral scalar fields the energy density and current, satisfying (\ref{g2}), are given by 
\begin{equation}
\vartheta_t = 
\frac{v}{4} : \left [(\der \phl)(\der \phl)(vt +x) + (\der \phr)(\der \phr)(vt -x) \right ]:\, , 
\label{endensity}
\end{equation}
and 
\begin{equation}
\vartheta_x = 
\frac{v^2}{4} :\left [(\der \phl)(\der \phl)(vt +x) - (\der \phr)(\der \phr)(vt -x) \right ]:\, ,
\label{encurrent}
\end{equation}
where $: \cdots :$ is the normal product in the oscillator algebras $\A_\pm$. In what follows we need the mean value 
of (\ref{encurrent}). By means of (\ref{bd},\ref{bd1}) one finds  
\begin{eqnarray} 
\langle \vt_x \rangle = 
\frac{\pi v^2}{12} \left (\frac{1}{\beta_2^2}-\frac{1}{\beta_1^2}\right ) + \qquad   
\nonumber \\
\frac{1}{4\pi} \left [ \left (\frac{\mu_2}{\zeta_+} - \frac{\mut_2}{\zeta_-}\right )^2 - 
\left (\frac{\mu_1}{\zeta_+} + \frac{\mut_1}{\zeta_-}\right )^2 \right ]\, . 
\label{t5}
\end{eqnarray}
Taking the expectation value of (\ref{g6}), we are ready at this point to derive the mean 
heat currents flowing through the gates $G_i$ in Fig. \ref{fig1}. Combining 
(\ref{t1},\ref{t4},\ref{t5}) one obtains in the gate $G_1$ 
\begin{eqnarray}
\langle q_x (G_1)\rangle = \frac{\pi v^2}{12} \left (\frac{1}{\beta_2^2}-\frac{1}{\beta_1^2}\right ) \qquad \quad  
\nonumber \\
+\frac{1}{2 \pi \zeta_+ \zeta_-}(\mu_1 \mut_2 + \mut_1 \mu_2 + \mu_1 \mut_1 -\mu_2 \mut_2)
\nonumber \\
+\frac{1}{4 \pi \zeta_+^2}(\mu_1-\mu_2)^2  + \frac{1}{4 \pi \zeta_-^2}(\mut_1-\mut_2)^2\, .
\label{t6}
\end{eqnarray}
Analogously, in the gate $G_2$ one has 
\begin{eqnarray}
\langle q_x (G_2)\rangle = \frac{\pi v^2}{12} \left (\frac{1}{\beta_2^2}-\frac{1}{\beta_1^2}\right ) \qquad \quad  
\nonumber \\
+\frac{1}{2 \pi \zeta_+ \zeta_-}(\mu_1 \mut_2 + \mut_1 \mu_2 + \mu_2 \mut_2 - \mu_1 \mut_1)
\nonumber \\
-\frac{1}{4 \pi \zeta_+^2}(\mu_1-\mu_2)^2  - \frac{1}{4 \pi \zeta_-^2}(\mut_1-\mut_2)^2\, .
\label{t7}
\end{eqnarray}
In analogy with the electric current (\ref{t2}), restricting (\ref{t6},\ref{t7}) at equilibrium (\ref{eq}), one obtains  
the persistent heat current 
\begin{equation}
\langle q_x (G_1)\rangle_{\rm eq} = \langle q_x (G_2)\rangle_{\rm eq} = \frac{\mu \, \mut}{\pi \zeta_+ \zeta_-} \, , 
\label{tpersistent}
\end{equation}
driven exclusively by both chemical potentials $\mu$ and $\mut$. 

In the next subsection we apply the above results for deriving 
the electric and heat conductance and compute the associated Lorenz number.

\subsection{Lorentz number and Wiedemann-Franz law} 

Using (\ref{t1}) one gets for the electric conductance in the gate $G_i$
\begin{equation}
E(G_i) = e^2 \frac{\der}{\der \mu_i} \langle j_x \rangle = (-1)^{i} \frac {e^2}{2 \pi \zeta_+^2} \, . 
\label{L1}
\end{equation} 
where the value $e$ of the electric charge has been restored. On the other hand, since 
\begin{equation}
\beta_i = \frac{1}{T_i k_{{}_{\rm B}}} \, , 
\label{T}
\end{equation}
$k_{{}_{\rm B}}$ being the Boltzmann constant, one obtains from  
(\ref{t6},\ref{t7})  
\begin{eqnarray}
H(G_i) &=& \frac{\der}{\der T_i} \langle q_x (G_i) \rangle  = -\beta_i^2 k_{{}_{\rm B}} \frac{\der}{\der \beta_i} \langle q_x (G_i) \rangle
\nonumber \\
&=& (-1)^{i} \frac {\pi v^2 k_{{}_{\rm B}}}{6 \beta_i} = 
 (-1)^{i} \frac {\pi v^2 k^2_{{}_{\rm B}}}{6} T_i\, , 
\label{L2}
\end{eqnarray} 
which is linear in the temperature $T_i$ as observed in \cite{KF-97}. 

In terms of (\ref{L1},\ref{L2}) the Lorentz number \cite{L-81} in the gate $G_i$ is 
\begin{equation}
L(G_i) = \frac{\beta_i k_{{}_{\rm B}} H(G_i)}{E(G_i)} = L_0\, v^2\, \zeta_+^2 \, , 
\label{L3}
\end{equation}
where 
\begin{equation}
L_0 = \frac{\pi^2}{3} \left (\frac{k_{{}_{\rm B}}}{e} \right )^2 \, , 
\label{L4}
\end{equation}
is the {\it free electron} value. As expected 
\begin{equation}
L(G_1) = L(G_2) \equiv L 
\label{LL}
\end{equation}
and using (\ref{z},\ref{v}) one finally gets 
\begin{equation}
L = L_0 \frac{(\k +g_-)^{1/2}(\k  +g_+)^{3/2}}{\k}\, , 
\label{L5}
\end{equation}
displaying explicitly the dependence on the statistical parameter $\k$ and the coupling constants $g_\pm$. 
We observe in this respect that 
\begin{equation}
L\vert_{{}_{g_\pm =0}} = \frac{\pi^2}{3} \left (\frac{k_{{}_{\rm B}}}{e} \right )^2 \k \equiv L^{\rm an}_0(\k) \, , 
\label{L6}
\end{equation}
represents the {\it free anyon} Lorenz number with statistical parameter $\k$. 
As expected, for canonical fermions one has $L^{\rm an}_0(1) = L_0$. Combining (\ref{L5}) and (\ref{L6}) one gets  
\begin{equation}
R\equiv \frac{L}{L^{\rm an}_0(\k)} = \frac{(\kappa +g_-)^{1/2}(\kappa +g_+)^{3/2}}{\kappa^2}\, , 
\label{L7}
\end{equation}
which codifies a temperature-independent deviation from the Wiedemann-Franz law \cite{WF-53} 
generated by the interaction. 
The vanishing of the Lorentz number $L$ at the boundary $\k = -g_\pm $ of the domain (\ref{physcond}) 
is a physical consequence of the vanishing of the velocity (\ref{v}) in these points.

\subsection{Mean entropy production and energy transmutation} 

We focus in this subsection on the entropy production, which represents the key quantity quantifying the departure from
equilibrium. In order to simplify the notation, we adopt here the modified chemical potentials 
\begin{equation}
\nu_i = \frac{\mu_i}{\zeta_+} \, , \qquad \nut_i = \frac{\mut_i}{\zeta_-} \, .
\label{nu}
\end{equation}
Plugging (\ref{t6}) and (\ref{t7}) in the general expression (\ref{g14}) one finds for the mean value of the entropy production  
\begin{eqnarray} 
\langle \dS \rangle = 
\frac{\pi v^2(\beta_1+\beta_2)(\beta_1-\beta_2)^2}{12\beta_1^2 \beta_2^2} + \quad 
\nonumber \\
\frac{\beta_1}{4\pi} B_1(\nu_i,\nut_i) + 
\frac{\beta_2}{4\pi}B_2(\nu_i,\nut_i)\, , 
\label{T2}
\end{eqnarray} 
where 
\begin{eqnarray} 
B_1(\nu_i,\nut_i)=(\nu_1-\nu_2-\nut_1+\nut_2)^2 + 4\nu_1\nut_1 \, ,  
\nonumber \\
B_2(\nu_i,\nut_i)=(\nu_1-\nu_2+\nut_1-\nut_2)^2 - 4\nu_2\nut_2\, .  
\label{T3}
\end{eqnarray} 
In order to implement the second law of thermodynamics we require that 
\begin{equation} 
\langle \dS \rangle \geq 0\, , \quad \forall \, \beta_i \geq 0\, , 
\label{T4}
\end{equation} 
which imposes some restriction on the chemical potentials $\{\nu_i,\nut_i\}$. In fact performing 
the repeated limits 
\begin{eqnarray}
\lim_{\beta_2\to \infty }\; \lim_{\beta_1 \to \infty} \frac{1}{\beta_1} \langle \dS \rangle = B_1(\nu_i,\nut_i)\, ,  
\label{t5a}\\
\lim_{\beta_1\to \infty }\; \lim_{\beta_2 \to \infty} \frac{1}{\beta_2} \langle \dS \rangle = B_2(\nu_i,\nut_i)\, , 
\label{5b}
\end{eqnarray}
we deduce that 
\begin{equation}
B_i(\nu_i,\nut_i) \geq 0 \, , \qquad i=1,2\, , 
\label{T6}
\end{equation}
are necessary conditions for the bound (\ref{T4}). From the explicit form (\ref{T2}) of $\langle \dS \rangle$ 
we infer that (\ref{T6}) are sufficient as well. 

The above considerations lead to the following conclusions: 
\medskip 

(a) The non-negativity (\ref{T4}) of the mean entropy production imposes 
non-trivial conditions (\ref{T6}) on the chemical potentials in the GGE heat reservoirs.  
\medskip 

(b) In the Gibbs limit in which one of the pairs $(\nut_1,\nut_2)$ or 
$(\nu_1,\nu_2)$ vanishes, one has that 
\begin{equation}
B_i(\nu_i,0) = (\nu_1-\nu_2)^2 \geq 0 \, , \quad B_i(0,\nut_i) = (\nut_1-\nut_2)^2 \geq 0\, , 
\label{T7}
\end{equation}
being identically satisfied. Therefore in that limit the entropy production is non-negative 
for any value of the chemical potentials. 
\medskip 

\begin{figure}[h]
\begin{center}
\begin{picture}(0,100)(120,20) 
\includegraphics[scale=0.36]{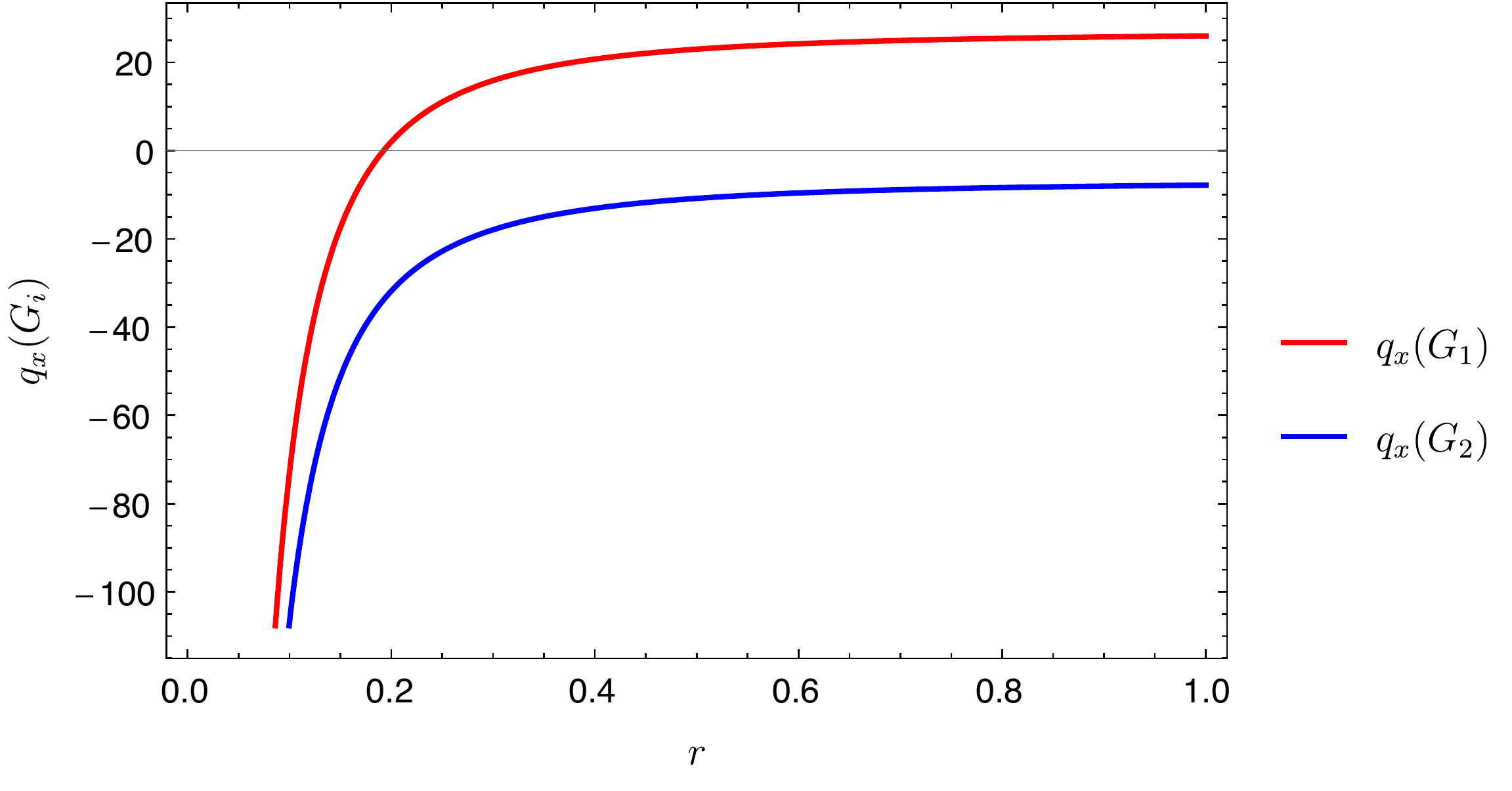}
\end{picture} 
\end{center}
\caption{(Color online) Heat flows (in units of $\beta$) through the hot and cold gates $G_1$ (red) and $G_2$ (blue) 
for a set of chemical potentials satisfying (\ref{T7}) .} 
\label{fig2}
\end{figure}

(c) The conditions (\ref{T6}) imply that 
\begin{equation} 
\langle \dQ \rangle = \frac{1}{4\pi}B_1(\nu_i,\nut_i) 
+\frac{1}{4\pi}B_2(\nu_i,\nut_i) \geq 0\, . 
\label{T8}
\end{equation}
where $\dQ$ is the observable defined by (\ref{g8}). Therefore in the physical regime (\ref{T7}) of non-negative 
mean entropy production our non-equilibrium anyon TL liquid converts chemical to heat energy without 
dissipation. Let us illustrate this aspect assuming without loss of generality that 
\begin{equation}
\beta_2 \geq \beta_1 \geq 0 \Longrightarrow r \equiv \frac{\beta_1}{\beta_2} \in [0,1]\, .    
\label{T9}
\end{equation}
Accordingly the hot and cold reservoirs in Fig. \ref{fig1} are respectively $R_1$ and $R_2$ because $T_1\geq T_2$. 
In this setting the heat flows in the gates $G_i$, where the leads $L_i$ are oriented 
as in Fig. \ref{fig1}, are shown in Fig. \ref{fig2}. We see that the heat current through the cold gate $G_2$ is always negative, 
indicating that the corresponding heat flow enters the cold reservoir. Concerning the heat current flowing in the hot gate $G_1$, 
there is a critical value $r_0$ (for the parameters chosen in 
Fig. \ref{fig2} one has $r_0 \sim 0.18$ ) for which $q_x(G_2)$ inverts his direction:  
for $0\leq r<r_0$ and $r_0 < r \leq 1$ it leaves and enters the hot reservoir respectively. 
Therefore, despite of the fact that the energy and particle currents have the same direction 
and intensity (see eqs. (\ref{g4}, \ref{g5})) in the gates $G_i$, this is not the case for the 
heat current because of the explicit dependence of $q_x(G_i)$ on the chemical potentials.   

\section{Anyon quantum noise} 

This section focusses on the quantum fluctuations described by the {\it connected} two-point electric current correlation 
function 
\begin{eqnarray}
\langle j_x(t_1,x_1) j_x(t_2,x_2)\rangle^{\rm con} = \qquad \qquad \quad 
\nonumber \\
\langle j_x(t_1,x_1) j_x(t_2,x_2)\rangle 
-\langle j_x(t_1,x_1)\rangle \langle j_x(t_2,x_2)\rangle \, . 
\label{no1}
\end{eqnarray}
Our main goal here is to investigate the dependence of these fluctuations on the 
statistical parameter $\kappa$, which opens the possibility to study experimentally 
the nature of the anyon TL excitations by measuring the noise. 

Using the definition (\ref{jx}) and (\ref{bd1}) one finds 
\begin{eqnarray}
\langle j_x(t_1,x_1) j_x(t_2,x_2)\rangle^{\rm con} = \qquad \qquad \quad 
\nonumber \\
-\frac{v^2}{4\zeta_+^2}\Biggl \{ \left [\beta_1\sinh \left (\frac{\pi}{\beta_1 }(vt_{12}- x_{12}) -
\ri \varepsilon \right )\right ]^{-2} + 
\nonumber \\
\left [\beta_2\sinh \left (\frac{\pi}{\beta_2 }(vt_{12} + x_{12}) -
\ri \varepsilon \right )\right ]^{-2} \Biggr \}\, , 
\nonumber \\
\label{no2}
\end{eqnarray} 
which shows that the second moment of the probability distribution generated by the current $j_x$ 
depends on the temperatures but does not involve the chemical potentials. 
The noise power at frequency $\omega$ is obtained \cite{BB-00} by performing the Fourier transform 
\begin{equation}
P(\omega) = \int_{-\infty}^{\infty} \e^{\ri \omega t} \langle j_x(t ,x) j_x(0,x )\rangle^{\rm con} \, . 
\label{no3}
\end{equation}
In what follows we use the temperatures $T_i$ defined by (\ref{T}) and set 
\begin{equation}
T_1 = T + \frac{\delta}{2}\, , \qquad T_2 = T - \frac{\delta}{2}\, . 
\label{no4}
\end{equation}
The integral in (\ref{no3}) can be performed explicitly and one finds in the limit $\varepsilon \to 0^+$ 
\begin{equation}
P(\omega) = 
\frac{\omega}{4 \pi \zeta_+^2} \left [ 2 + \coth \left (\frac{\omega}{2 T v - v \delta } \right ) + 
\coth \left (\frac{\omega}{2 T v + v \delta} \right ) \right ]\, ,  
\label{no5}
\end{equation}
which is the subject of the analysis below. 

\begin{figure}[h]
\begin{center}
\begin{picture}(0,100)(120,20) 
\includegraphics[scale=0.36]{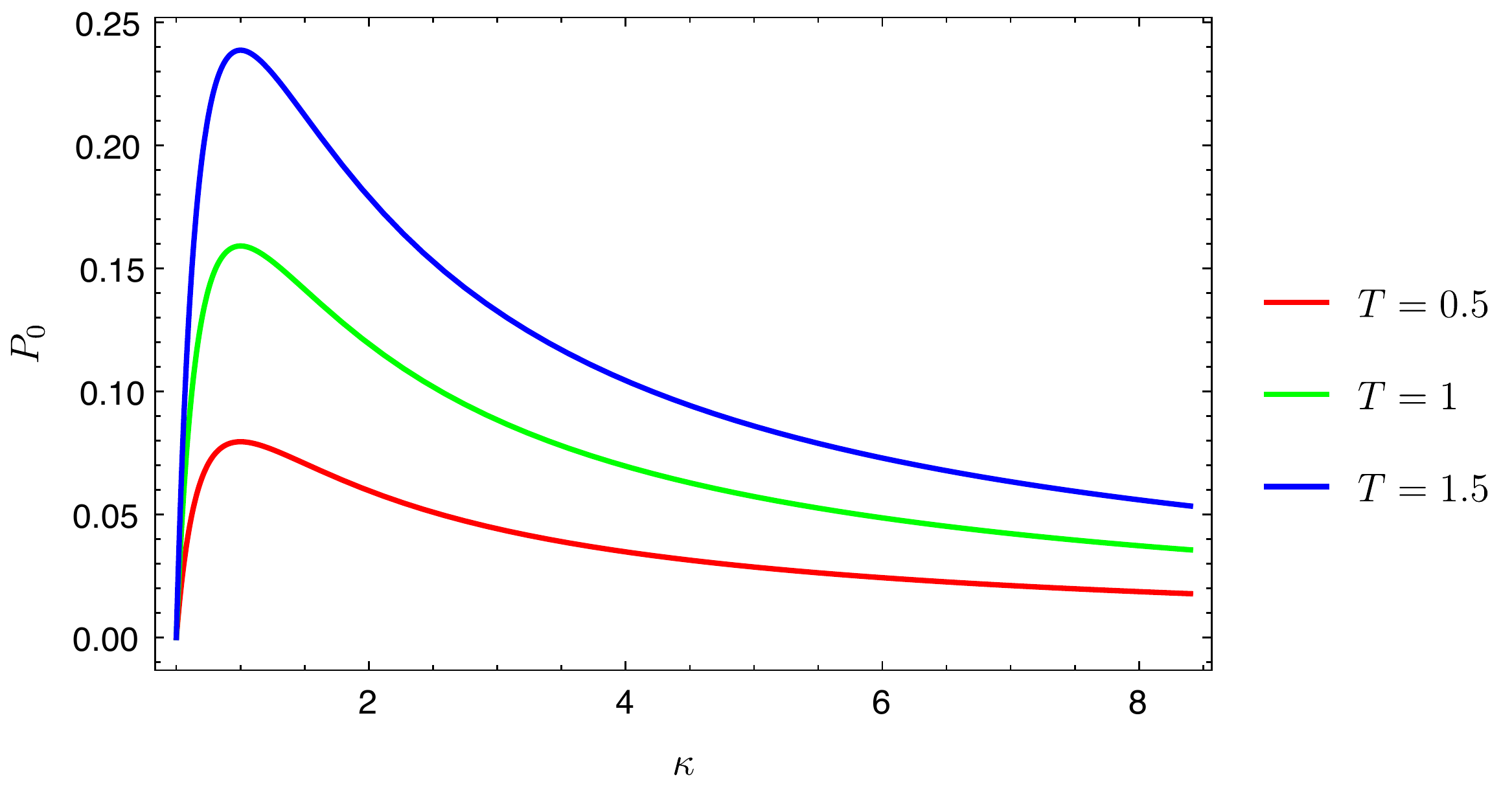}
\end{picture} 
\end{center}
\caption{(Color online) Dependence of $P_0$ on $\kappa$ for $g_-=-0.5$ and three different values of the temperature.} 
\label{fig3}
\end{figure}

\begin{figure}[h]
\begin{center}
\begin{picture}(0,100)(120,20) 
\includegraphics[scale=0.36]{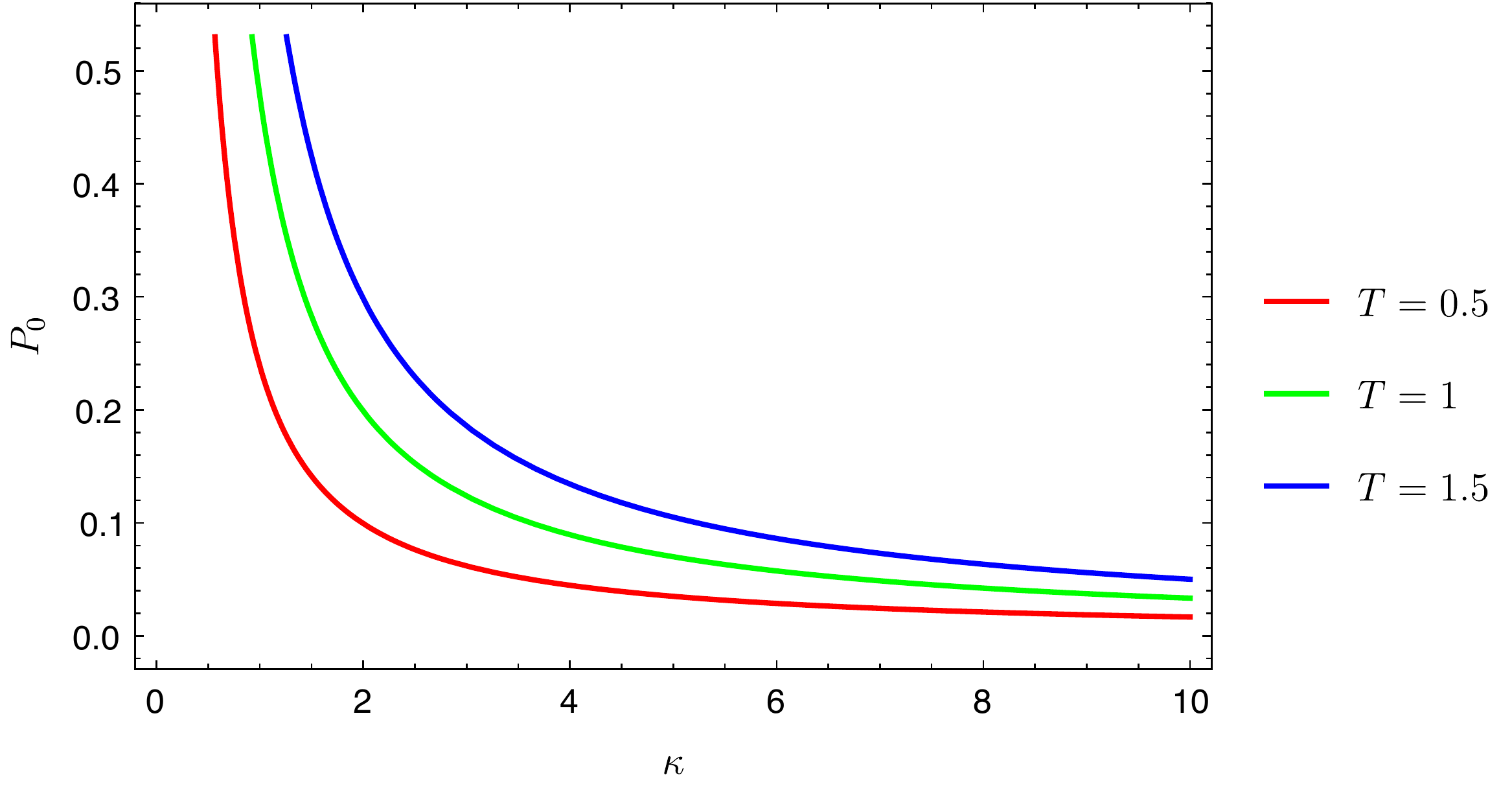}
\end{picture} 
\end{center}
\caption{(Color online) Dependence of $P_0$ on $\kappa$ for $g_-=0.5$ and three different values of the temperature.} 
\label{fig4}
\end{figure}

First of all, the zero-frequency limit in (\ref{no5}) gives 
\begin{equation}
P_0 = \lim_{\omega \to 0} P(\omega) = \frac{v}{\pi \zeta_+^2} T = 
\frac{(\kappa + g_-)}{\pi \kappa^2} T \, , 
\label{no6}
\end{equation}
which shows the typical linear in the temperature Johnson-Nyquist behaviour. 
The interesting feature is the dependence on the statistical parameter 
$\kappa \in \cal D$ (see eq. (\ref{physcond})). There are two characteristic regimes 
depending on the sign of coupling constant $g_-$. For $g_- < 0$ the admissible values of $\kappa $ 
are $\kappa \geq -g_-$ and the coefficient in (\ref{no6}) has a maximum at $\kappa = -2 g_-$. 
In this case the behaviour of $P_0$ 
is shown in Fig. \ref{fig3} for three different values of the temperature. 
For $g_-\geq 0$ the allowed values for $\kappa$ are $\kappa >0$ and the noise 
$P_0$ is monotonically decreasing as shown in Fig. \ref{fig4}. 
In both cases $P_0$ decays as $1/\kappa$ for large $\kappa$. 

\begin{figure}[h]
\begin{center}
\begin{picture}(0,100)(120,20) 
\includegraphics[scale=0.36]{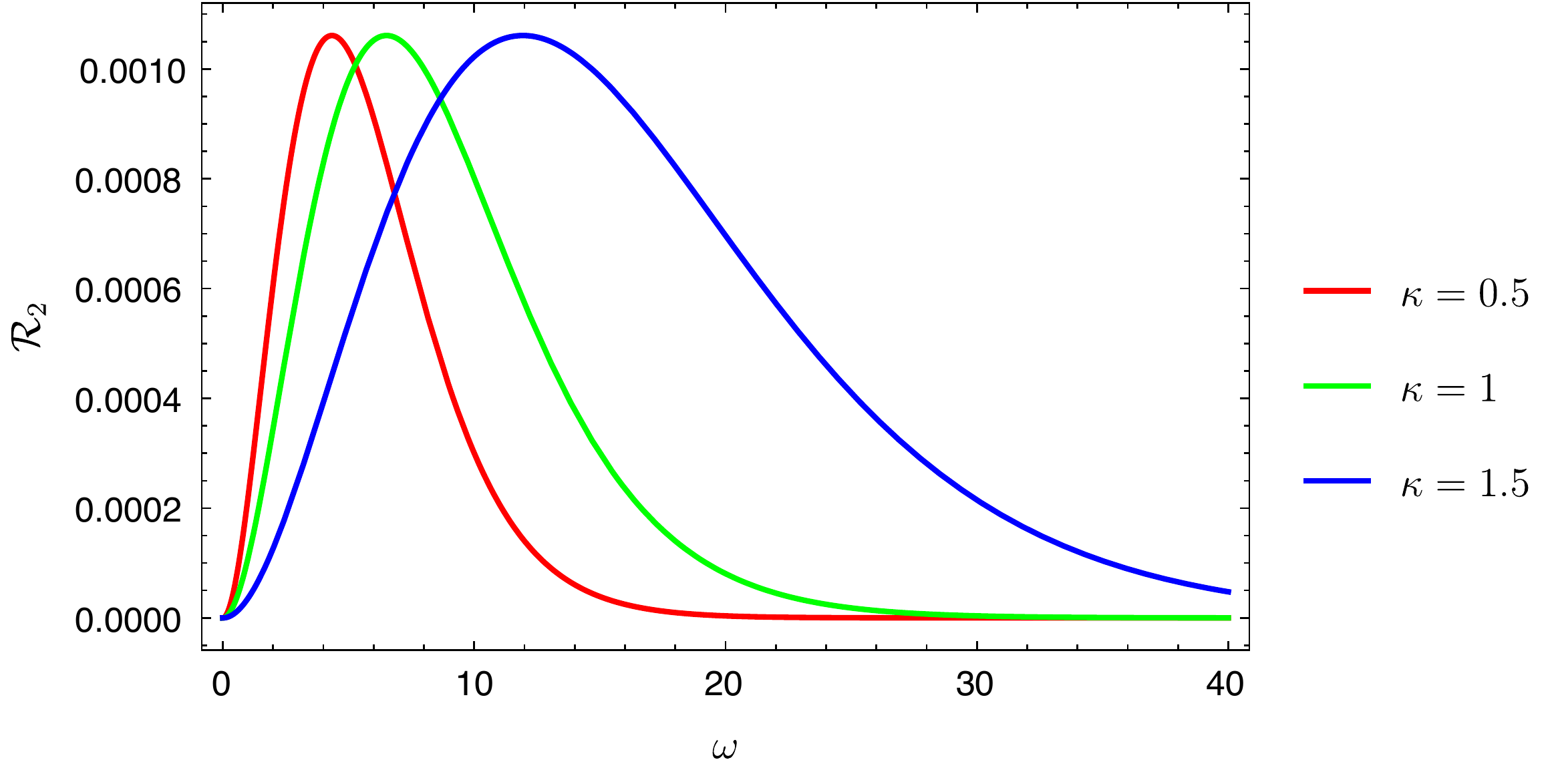}
\end{picture} 
\end{center}
\caption{(Color online) Dependence of ${\cal R}_2$ on $\omega$ for three different values of $\kappa$ and $g_+=g_-=-0.5$.} 
\label{fig5}
\end{figure}

\begin{figure}[h]
\begin{center}
\begin{picture}(0,100)(120,20) 
\includegraphics[scale=0.36]{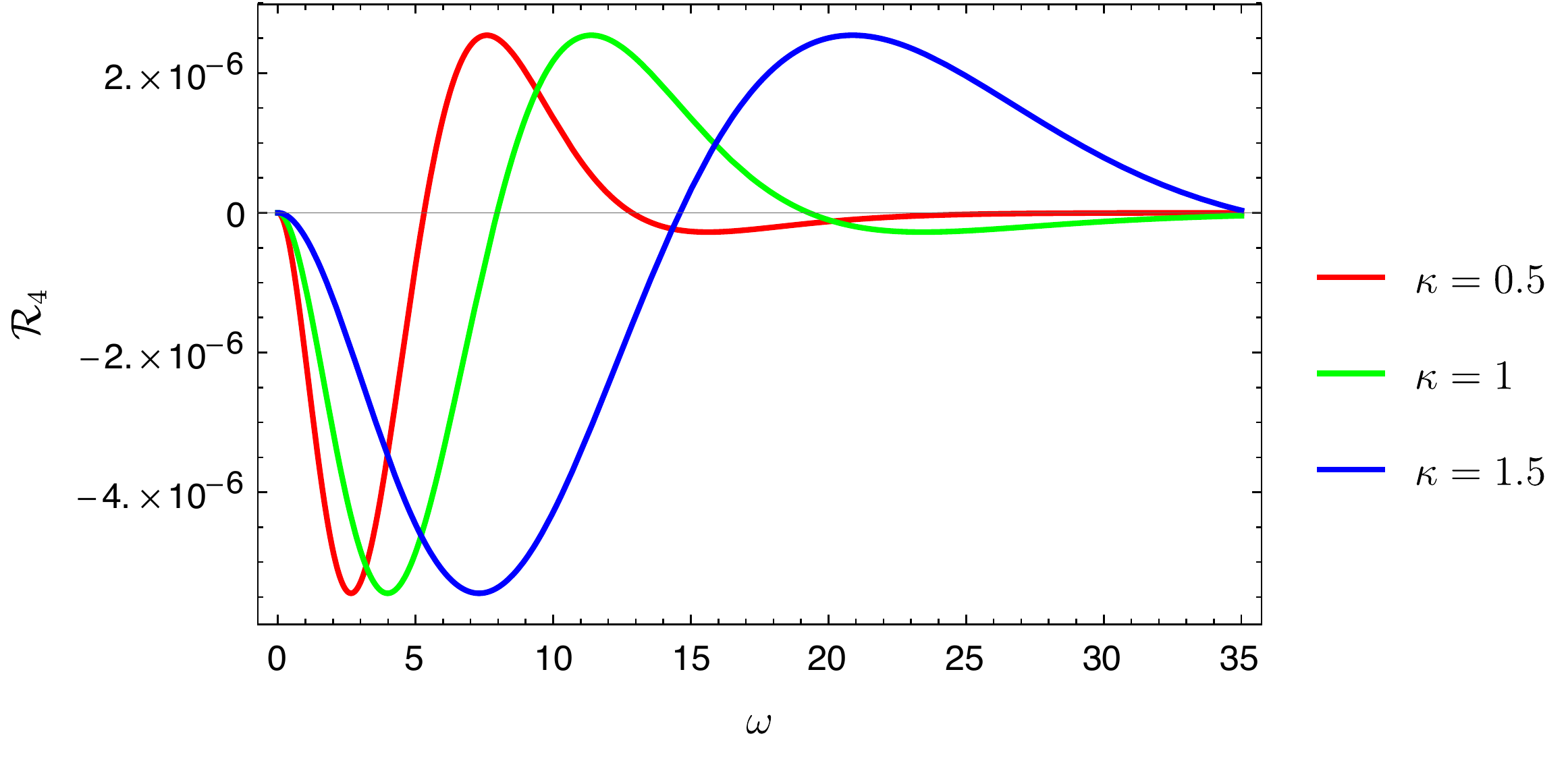}
\end{picture} 
\end{center}
\caption{(Color online) Dependence of ${\cal R}_4$ on $\omega$ for three different values of $\kappa$ and $g_+=g_-=-0.5$.} 
\label{fig6}
\end{figure}

Let us explore finally the frequency dependence of $P(\omega)$. For this purpose we expand (\ref{no5}) around $\delta =0$.  
One has 
\begin{equation} 
P(\omega ) =  Q(\omega) \left [ 1 + {\cal R}_2(\omega)\, \delta^2 + {\cal R}_4(\omega)\, \delta^4+... \right ]\, , 
\label{no7}
\end{equation}
where 
\begin{equation}
Q(\omega) = \frac{\omega}{2 \pi \zeta_+^2} \left [1 + \coth \left (\frac{\omega}{2 T v)} \right )\right ]\, , 
\label{no8}
\end{equation}
\begin{equation}
{\cal R}_2(\omega) = \frac{\omega}{16 T^4 v^2} 
\frac{\left [\omega \coth \left (\frac{\omega}{2 T v} \right )-2Tv \right ]}
{\left [\omega \coth \left (\frac{\omega}{2 T v} \right )+1\right ] \sinh^2 \left (\frac{\omega}{2 T v} \right )}\, , 
\label{no9}
\end{equation}
and a similar but longer expression for ${\cal R}_4$, whose explicit form is not reported for conciseness. 
The $\kappa$-dependence of the coefficients ${\cal R}_i$ is carried by the velocity $v$ given by (\ref{v}). 
Figs. \ref{fig5} and \ref{fig6} illustrate the impact of $\kappa$ on the frequency behaviour. The 
frequencies where ${\cal R}_2$ reaches his maximum and ${\cal R}_4$ his minimum and maximum 
manifestly depend on $\kappa$. Therefore the frequency behaviour of the noise is sensitive to 
the specific value of the statistical parameter of the anyon excitations which are propagated. 

The noise generated by the helical current $\jt_x$ can be investigated along the above lines as well. 
As a consequence of the axial-vector duality, in this case the power ${\widetilde P}(\omega)$ is simply 
obtained from (\ref{no5}) by the substitution $\zeta_+ \longmapsto \zeta_-$. 

In conclusion, both the zero and finite frequency current quantum fluctuations carry the imprint of the 
anyon statistics and offer therefore relevant experimental tools for detecting the statistical parameter $\k$.

\section{Discussion} 

We performed a systematic study of the dependence of the anyon particle and heat transport on the 
statistical parameter $\k$ of the TL anyon liquid in contact with two GGE heat reservoirs. Each of them 
depends on two chemical potentials, which implement the axial-vector duality of the model. 
The system is a specific case of non-equilibrium CFT with central charge $c=1$ and provides 
an instructive example for testing some general ideas in this context. 

In this setting we derived in explicit form the mean value of the particle and heat currents 
and determined the associated Lorentz number, which shows a $\k$-dependent 
deviation from the Wiedemann-Franz law. 
We also investigated the mean value of the entropy production $\langle \dS \rangle$, 
generated by the spontaneous breaking of time reversal,  
and established the conditions on the chemical potentials implementing the physical requirement 
$\langle \dS \rangle \geq 0$. We observed that the mean values of the electric and helical currents 
depend on the chemical potentials but not on the temperatures. Precisely the opposite behaviour 
is characterising instead the associated quantum fluctuations, which is consistent with the 
general CFT predictions \cite{BD-15, HL-18}. We have shown in addition that the quadratic 
fluctuations of the anyon electric current in the zero frequency limit obey the Johnson-Nyquist 
law with $\k$-dependent pre-factor. The noise at finite frequencies carries specific 
$\k$-dependence as well, providing attractive experimental applications. 

The framework, developed in this paper, can be applied in different contexts and 
generalised in various directions. An attractive subject in the context of CFT current algebras \cite{GO-86} is  
the detailed study of the non-equilibrium representation of the axial-vector current algebra generated in 
the TL model. Moreover, it would be interesting to extend the results of this paper to other types 
of anyon quantum liquids, which have been considered in the literature \cite{CM-07}-\cite{P-19}. 
Another challenging issue is to explore along the lines of \cite{MSS-17, MSS-18} the behaviour 
of the higher moments of the probability distribution generated by the entropy production operator. 
One will obtain in this way a complete picture of the departure from equilibrium, induced by the contact with 
the two GGE reservoirs. We will come back to these issues elsewhere.

\bigskip
\leftline{\bf Acknowledgments} 
\medskip 

M. M. and P. S. would like to thank the Laboratoire d'Annecy-le-Vieux de Physique Th\'eorique 
and INFN, Sezione di Pisa respectively, for the kind hospitality at the early stage of this investigation. 







\end{document}